# Breakthroughs in Photonics 2014: Relaxed Total Internal Reflection

**Saman Jahani and Zubin Jacob**

(Invited Paper)

Department of Electrical and Computer Engineering, University of Alberta,
Edmonton, Alberta, T6G 2V4, Canada

Corresponding author: Z. Jacob (e-mail: zjacob@ualberta.ca).

**Abstract**: Total internal reflection (TIR) is a ubiquitous phenomenon used in photonic devices ranging from waveguides and resonators to lasers and optical sensors. Controlling this phenomenon and light confinement are keys to the future integration of nanoelectronics and nanophotonics on the same silicon platform. We introduced the concept of relaxed total internal reflection in 2014 to control evanescent waves generated during TIR. These unchecked evanescent waves are the fundamental reason photonic devices are inevitably diffraction-limited and cannot be miniaturized. Our key design concept is the engineered anisotropy of the medium into which the evanescent wave extends thus allowing for skin depth engineering without any metallic components. In this article, we give an overview of our approach and compare it to key classes of photonic devices such as plasmonic waveguides, photonic crystal waveguides and slot waveguides. We show how our work can overcome a long standing issue in photonics – nanoscale light confinement with fully transparent dielectric media.

## 1. Introduction

In 1611, Johannes Kepler, a German mathematician and astronomer, discovered the phenomenon of total internal reflection ten years before Willebrord Snell derived his famous formula for the refraction of light. It is interesting to note that 200 years lapsed before it was accidentally discovered by Daniel Colladon in 1842 that TIR can be used for light guiding. TIR is the fundamental operating principle behind many photonic devices, such as optical dielectric waveguides [1], total internal reflection fluorescence microscopes [2] and laser cavities [3].

The emergence of nanoscale photonics led to revisiting the laws of refraction and reflection through artificially structured materials, such as chiral materials [4], negative index materials [5], [6], photonic crystals [7], and bio-inspired structures [8], [9]. Many interesting phenomena were observed, which are not seen in conventional dielectrics. Recent work has also shown the existence of a generalized snell's law to control refraction and reflection of light using metasurfaces which control the phase gradient at the interface between dielectric media [10], [11]. This allows the control of an optical ray with arbitrary phase and amplitude [12], [13].

However, all of these approaches dealt primarily with propagating waves. In 2014, we introduced relaxed total internal reflection to control the evanescent waves produced during total internal reflection [14], [15]. The skin depth of evanescent waves at total internal reflection is the fundamental reason dielectric photonics is diffraction-limited in size. We have shown that relaxed total internal reflection requires unique anisotropy and is achievable using lossless semiconductor building blocks. This can surprisingly lead to sub-diffraction light confinement in photonic waveguides (no metallic components).

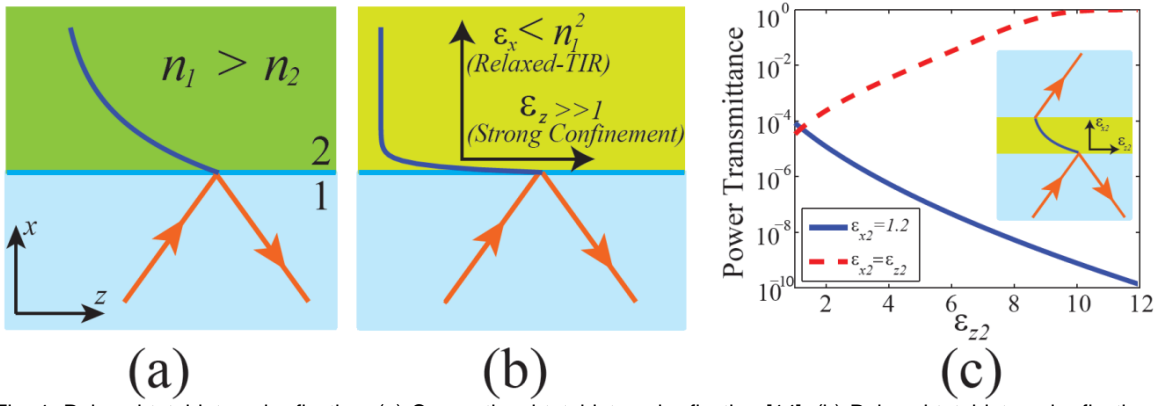

Fig. 1. Relaxed total internal reflection. (a) Conventional total internal reflection [14]. (b) Relaxed total internal reflection [14]. (c) Controlling the evanescent wave tunneling of a p-polarized light using transparent anisotropic metamaterials. In all cases, $\varepsilon_y = \varepsilon_z$. Fundamentally different scaling of tunneled energy is observed for the two cases.

## 2. Relaxed Total Internal Reflection

Conventional total internal reflection occurs when a ray from a medium with a greater refractive index crosses the interface of a medium with a lower refractive index ($n_1 > n_2$) and the incident angle is greater than the so-called critical angle ($\theta_c = \sin^{-1}(n_2/n_1)$). Under these conditions the ray is totally reflected to the first medium and decays evanescently inside the second medium [See Fig. 1 (a)]. These conditions hold true for both s and p polarized incidence. However, we have recently shown that if the second medium is a uniaxial dielectric with optical axis normal to the interface ($\varepsilon_z = \varepsilon_y \neq \varepsilon_x$), the TIR condition for s and p polarizations are separated to [14]:

$$\begin{cases} n_1 > \sqrt{\varepsilon_y}, & s-polarization \\ n_1 > \sqrt{\varepsilon_x}, & p-polarization \end{cases} \quad (1)$$

We call this phenomenon as relaxed total internal reflection since only one component needs to be smaller than the high index dielectric. This fundamentally leaves open another degree of freedom for engineering the skin depth of evanescent waves. As mentioned before, this skin depth of evanescent waves has gone surprisingly overlooked in dielectric photonic devices.

For p-polarization, there is one more degree of freedom – the dielectric component ($\varepsilon_z$) for controlling the evanescent wave in the second medium. We showed that if $\varepsilon_z \gg 1$, the skin depth in the second medium can be reduced by a factor of $\sqrt{\varepsilon_z/\varepsilon_x}$ [See Fig. 1 (a)] [14], [15]. Note that this strong anisotropy does not exist in natural dielectrics, but it is achievable using engineered nanostructures [16]–[18]. It can easily be shown that the relaxed-TIR and strong confinement conditions can be applied even if the anisotropic medium is biaxial ($\varepsilon_y \neq \varepsilon_z$). In this case, the relaxed-TIR condition for p-polarization is still $n_1 > \sqrt{\varepsilon_x}$, but for s-polarization is $n_1 > \max\{\sqrt{\varepsilon_y}, \sqrt{\varepsilon_z}\}$. To obtain strong confinement of the evanescent wave in the second medium for p-polarization, we should have $\varepsilon_y, \varepsilon_z \gg 1$.

Fig. 1 (c) displays tunneling of evanescent waves through a transparent anisotropic thin slab (solid line) compared to an isotropic slab (dashed line). They show fundamentally different behavior with increasing dielectric constant perpendicular to the interface. The slab thickness is $1\lambda$ and it is sandwiched between two infinite silicon ($\varepsilon = 12$) half spaces. The incident angle is $\theta_{in} = 60°$. When the permittivity of the isotropic dielectric increases, skin depth inside the slab increases. This causes the transmission to increase since tunneling increases with increasing skin depth. Note that when $\varepsilon_2 > 9$, the incident angle is lower than the critical angle. Thus the transmission goes up rapidly. On the other hand, if the slab is anisotropic and permittivity increases only in the $z$ direction, the skin depth reduces due to the strong confinement of the evanescent wave. This explains the drop in transmission with increase of index. This control over tunneling using anisotropy opens fundamentally new avenues for nanophotonics.

## 3. Transparent Sub-diffraction Photonics

Light has a characteristic size set by its wavelength which is around a micron at optical communication frequencies and fundamentally impedes any effort to integrate it on current industrial standards of nanoscale circuitry. As micron scale fibers and waveguides are made smaller, the fundamental roadblock arises since light escapes out of the core of the fiber even if the core is made from a high refractive index dielectric such as silicon [19]. This causes cross talk between adjacent waveguides in dense photonic integrated circuits.

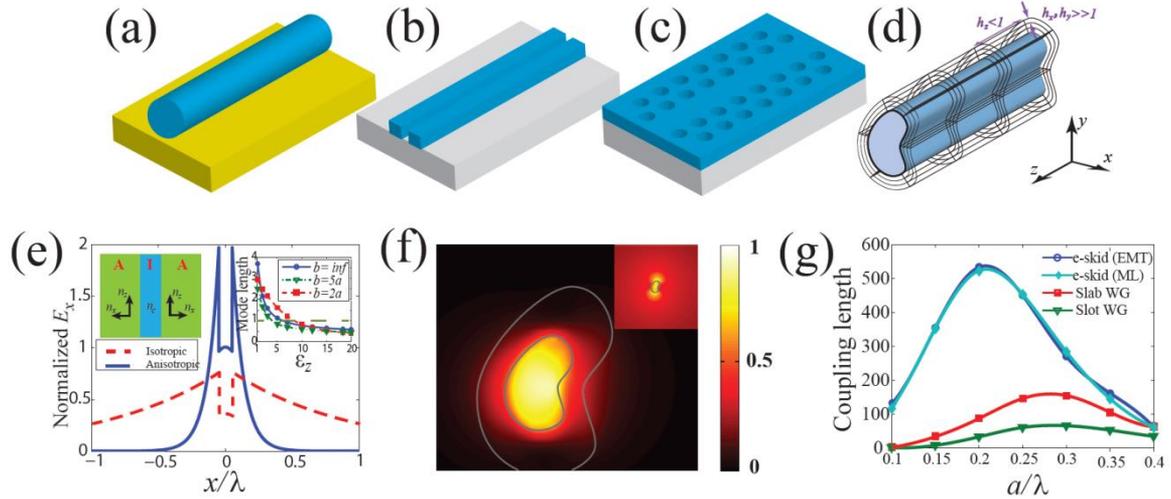

Fig. 2. Various approaches to nano-waveguiding. (a) Hybrid plasmonic waveguide. (b) Dielectric slot waveguide. (c) Photonic crystal waveguide. (d) Extreme skin depth (e-skid) waveguide [15]. (e) Sub-diffraction confinement in 1D e-skid waveguide [14]. (f) Strong confinement in 2D e-skid waveguide [14]. (g) Reducing cross talk between two coupled e-skid waveguides [14].

The search for replacement of optical fibers in photonic integrated circuits has led to the design of many photonic waveguide architectures. These waveguides can be categorized in two major classes: metallic (or plasmonic) and dielectric waveguides. Metallic waveguides are predominantly used for guiding electromagnetic waves at microwave frequencies due to the specular reflection of metals at low frequencies. However, due to the considerable skin depth of metals at optical frequencies, it is difficult to scale microwave waveguides for higher frequency applications. A number of architectures have recently emerged to effectively utilize metals for waveguiding at optical frequencies. These include the long range surface plasmon polariton (SPP) on metal strip waveguides (IMI, I≡insulator, M≡metal) [20] which are useful for sensing applications but not for sub-diffraction confinement. The inverse design consisting of MIM waveguides [21], confines light to subwavelength scales but it leads to low propagation lengths. V-groove [22] and wedge [23] plasmons are excellent candidates for relatively long range propagation and sub-diffraction confinement however excitation and detection of these modes as well as interfacing with existing silicon waveguide technology are major challenges. Recently, hybrid dielectric-plasmonic waveguides [See Fig. 2 (a)] [24] have emerged that confine light in a low index gap above metals reducing the field penetration in the metal thus allowing for increased propagation length. Another alternative is an epsilon-near-zero metamaterial waveguide [25] which allows modes to tunnel through subwavelength size structures. However, due to the optical losses of metals at optical frequencies, the propagation loss in the above mentioned sub-diffraction plasmonic structures is relatively high [23].

Dielectric slot-waveguides [See Fig. 2 (b)] and photonic crystal waveguides [See Fig. 2 (c)] are the two major all-dielectric architectures which are able to strongly confine light. In slot-waveguides, the electric field intensity goes up in a tiny low index gap surrounded by two high index rods. This allows sub-diffraction confinement of the mode inside the gap [26]. Photonic crystal waveguides can contain and confine light in defects in the designed photonic band gap. If there is no disorder, light cannot escape from the core even at sharp bends [27]. However, in both of these structures, if two waveguides are fabricated close to each other, light can easily be coupled to the adjacent waveguide. Thus it causes considerable cross talk in photonic integrated circuits [28].

We recently proposed a new class of transparent anisotropic claddings to reduce skin depth using relaxed-TIR rules [14], [15]. The skin depth reduction in the cladding can lead to sub-diffraction confinement of the mode inside the so-called extreme skin depth (e-skid) waveguide [See Fig. 2 (d)] [14]. Fig. 2 (e) displays the electric field distribution of the first TM mode of a 1D e-skid waveguide in comparison with conventional slab waveguides [14]. The core is glass with a thickness of $0.1\lambda$. The anisotropic metamaterial has relative permittivity of $\varepsilon_x = 1.2$ and $\varepsilon_z = 15$. It is seen that if $\varepsilon_z > 7$, sub-diffraction confinement is achievable. This is contrary to years of waveguide literature which claimed that transparent media cannot confine light to below the diffraction limit.

We can implement this transparent anisotropic metamaterial to cover an arbitrarily shaped 2D waveguide [15]. Electric energy density for the bare waveguide and e-skid waveguide is displayed in fig. 2 (f) [14]. The core is glass with an average radius of $0.1\lambda$. The anisotropic cladding has a relative permittivity of $\varepsilon_x = 1.2$ and $\varepsilon_z = 15$. Without the cladding, less than 1% of the energy is confined inside the core, but using the transparent anisotropic cladding, 36% of the total power is confined inside the glass core and the mode area is reduces from $80A_0$ to $0.7A_0$ ( $A_0 = (\lambda/2n_{core})^2$ ).

The major advantage of e-skid waveguide for practical application is reduction of cross talk at telecommunication wavelength ($\lambda = 1550 nm$). Fig. 2 (g) shows the coupling length between two coupled silicon slab waveguides with a thickness of $0.1\lambda$ and center-to-center separation of $0.5\lambda$ [14]. The anisotropic cladding is composed of a Ge/SiO$_2$ multilayer metamaterial with Ge filling fraction of $\rho = 0.6$. The metamaterial has an effective permittivity of $\varepsilon_x = 4.8$ and $\varepsilon_z = 11.9$. It is clearly seen that the coupling length in e-skid waveguides is an order of magnitude larger than that in conventional slab waveguides. The coupling length for slot-waveguides is comparable to or lesser than that for conventional slab waveguides. This makes e-skid waveguides suitable for dense photonic integration applications.

## 4. Conclusion

We presented a review on relaxed total internal reflection and strong confinement conditions. We showed that the evanescent wave tunneling can be controlled using anisotropic metamaterials. This leads to extremely low skin depth in metamaterials, which is suitable for transparent sub-diffraction confinement in photonic waveguides. The proposed waveguides can be used to increase photonic integration density due to the reduction in cross talk between closely packed waveguides.